\documentclass[12pt]{article}
\usepackage{epsf}
\usepackage{graphicx}
\usepackage{lscape}
\usepackage{amssymb}
\usepackage{pazh}
\tightenlines

\def\ΧΆ{$\pm$}
\def\*{$^{*}$}

\def\a{$^{\mbox{\small a}}$}
\def\b{$^{\mbox{\small b}}$}
\def\c{$^{\mbox{\small c}}$}
\def\d{$^{\mbox{\small d}}$}

\def\Œ³ŒÀ{\mbox{\<<Œ³ŒàŒÐŒÝŒÐŒâ\>>}}

\def\deg{$^\circ$}
\def\ŒÕŒàŒÓŒá{ŒíŒàŒÓ~Œá$^{-1}$}
\def\ŒÕŒàŒÓŒáŒÜ{ŒíŒàŒÓ~ŒáŒÜ$^{-2}$~Œá$^{-1}$}

\begin{document}
 
{\footnotesize Astronomy Letters, Vol. 31, No. 6, 2005, pp. 380-387. 
Translated from Pis'ma v Astronomicheskii Zhurnal, Vol. 31, No. 6, 
2005, pp. 427-436.
Original Russian Text Copyright \copyright\, 2005 by Tsygankov, 
Lutovinov.}

%\hline
%\hline

\title{\bf Long-Term INTEGRAL and RXTE Observations of the X-Ray Pulsar LMC X-4}   
 
\author{\bf \hspace{-1.3cm}\copyright\, 2005 \ \
   S.S.Tsygankov\affilmark{1}$^{\,*}$, A.A.Lutovinov\affilmark{1}}     
 
\affil{
$^1$ {\it Space Research Institute, Russian Academy of Sciences, 
Profsoyuznaya ul. 84/32, Moscow 117810, Russia}}
 
\vspace{2mm}
%\received{~~~~~~~~}

\sloppypar 
\vspace{2mm}
\noindent

%\abstract

{  
We analyze the observations of the X-ray pulsar LMC X-4 performed by the INTEGRAL
observatory and the All-Sky Monitor (ASM) of the RXTE observatory over a wide energy range. The
observed hard X-ray flux from the source is shown to change by more than a factor of 50 (from  $\sim70$ mCrab
in the high state to  $\sim1.3$ mCrab in the low state) on the time scale of the accretion-disk precession period,
whose mean value for 1996--2004 was determined with a high accuracy, $P_{prec} = 30.275 \pm 0.004$ days. In
the low state, a flare about 10 h in duration was detected from the source; the flux from the source increased
by more than a factor of 4 during this flare. The shape of the pulsar's broadband spectrum is essentially
invariable with its intensity; no statistically significant features associated with the possible resonance
cyclotron absorption line were found in the spectrum of the source. 
\copyright\, 2005 Pleiades Publishing Inc.}


\vfill
 
{$^{*}$ E-mail: st@hea.iki.rssi.ru}
\newpage
\thispagestyle{empty}
\setcounter{page}{1}

\section*{INTRODUCTION}

    The X-ray pulsar LMC X-4 in the Large Magellanic 
Cloud (LMC) (the distance to the object is $d=50$
kpc) is a high-mass binary with a pulsation period
of  $\sim13.5$ s (Kelley et al. 1983), in which the compact
object is eclipsed by its optical companion, an O8
star of the 14th magnitude with a mass of 20 $M_{\odot}$
(Chevalier and Ilovaisky 1977), every  $\sim$1.408 days
(Li et al. 1978; Lang et al. 1981; Levine et al. 2000).
Levine et al. (2000) provided the following orbital
parameters for the binary: $a_{x}$sin$i=26.333\pm0.019$ light seconds,
the eccentricity $e<0.003 (2\sigma)$, and the epoch of the
zero orbital phase $T_{0}=$MJD$51110.86571$.
  
  Lang et al. (1981) pointed to the existence of a
superorbital period in the binary,  $\sim30.5$ days, within
which the intensity of the source changes by a factor
of about 60. These authors also assumed that this
effect is produced by the blockage of direct X-rays by
a precessing accretion disk tilted with respect to the
orbital plane, much as is the case in the binary Her X-1 
(Tananbaum et al. 1972). Based on the measurements 
of almost the same X-ray flux observed from
the source during eclipses in its low and high states,
Woo et al. (1995) concluded that the intrinsic luminosity 
of the pulsar is constant, which also confirms
the assumption made by Lang et al. (1981). Having
analyzed the RXTE and GINGA observations, Paul
and Kitamoto (2002) estimated the rate of decrease
in the precession period of the accretion disk to be
$\dot P\sim-2\times10^{-5}$ s s$^{-1}$.
 
    The flaring activity of LMC X-4 has been widely
discussed in the literature (see, e.g., Epstein et
al. 1977; Skinner et al. 1980; Kelley et al. 1983;
Levine et al. 1991). A flare with a duration from
dozens of seconds to dozens of minutes is detected
from the source during its high state, on average,
once a day. However, there is also evidence (Woo et
al. 1995) for the presence of bursts in the low state.

    In different periods, the pulsar LMC X-4 demonstrates 
both a spin-down and a spin-up, suggesting
that its period is close to its equilibrium value. In this
case, according to the disk-accretion theory (Ghosh
and Lamb 1979), a neutron star with an equilibrium
period of  $\sim13.5$ s and an X-ray luminosity of 
$\sim 4\times10^{38}$ erg s$^{-1}$ would have a very high magnetic
moment,  $\sim10^{31.5}$ G cm$^{3}$ (Woo et al. 1996).

    The spectrum of the source above several keV is
described by a power law with a high-energy cutoff.
The measured absorption column density toward the
source (along with the best-fit parameters) does not
depend on the phase of the precession period; its
value, $N_H\sim5.5\times10^{20}$ atoms cm$^{-2}$, is close to the
Galactic column density, suggesting the absence of
strong internal absorption in the binary (Naik and
Paul 2003). Different authors (Levine et al. 1991; Mihara 
1995; Woo et al. 1996; La Barbera et al. 2001)
obtained several estimates for the presence of a cyclotron 
absorption line in the source's spectrum in the
energy range 19 to 100 keV, but as yet there are no reliable 
measurements of its intensity and, accordingly,
the surface magnetic field of the neutron star.

    In this paper, we present the results of our timing 
and spectral analyses for the pulsar LMC X-4
performed by using INTEGRAL and RXTE observations 
over a wide (1--100 keV) energy range. Preliminary 
results were obtained by Lutovinov et al. (2004).

\section*{OBSERVATIONS}

    The INTEGRAL International observatory (Winkler 
et al. 2003), which carries four scientific instruments 
that allow the emission from astrophysical
objects to be studied over a wide wavelength range
(from optical to hard $\gamma$ rays), was placed in orbit
by a Russian Proton launcher on October 17, 2002
(Eismont et al. 2003).
  
  In this paper, we analyze the data obtained with
the ISGRI detector of the IBIS gamma-ray telescope 
(Ubertini et al. 2003) and the JEM-X X-ray 
monitor of the INTEGRAL observatory (Lund
et al. 2003). The ISGRI detector has an effective
area of  $\sim960$ cm$^2$ at 50 keV, is effectively sensitive
to photons in the energy range 20 to 200 keV (the
energy resolution is  $\sim$7\% at 100 keV), and allows
the image of the sky within a 29$^{\circ}\times$29$^{\circ}$ field of view
(the full coding zone is 9$^{\circ}\times$9$^{\circ}$) to be reconstructed
with a nominal spatial resolution of  $\sim12$ arcmin (the
angular size of the mask element). See Lebrun et
al. (2003) for a more detailed description of the detector. 
The main elements of the two identical JEM-X 
modules (JEM-X1 and JEM-X2) are coded-mask
xenon­methane gas chambers placed at about 3.2 m
from the plane of the detectors. Each module of the
telescope has the following technical characteristics:
the energy range 3--35 keV, the field of view (the
full coding zone) 13.2\deg (4.8\deg) in diameter, and the
detector area 500 cm$^2$ (Lund et al. 2003).

    About 500 individual pointings were made as
part of the studies of the LMC region performed
by the INTEGRAL observatory from January 2
through January 28, 2003, during which the pulsar 
LMC X-4 was within the field of view of the
INTEGRAL instruments. The total observing time
for the source was  $\sim10^{6}$ s; at all pointings, it was
in the full coding zone. We also used data from the
All-Sky Monitor (ASM) of the RXTE observatory
(http://xte.mit.edu/ASM$_{-}$lc.html) to analyze the
long-period characteristics of the source. These data
are daily averaged fluxes from the source in the energy
range 1.3--12.2 keV.

    We constructed an X-ray image and performed a
spectral analysis of the IBIS data using the methods
described by Revnivtsev et al. (2004) and Lutovinov
et al. (2003). An analysis of a large set of calibration
observations for the Crab Nebula revealed a systematic 
error of  $\sim$10\% in the estimated absolute flux
from the source over a wide energy range; the spectral
shape was reconstructed with an accuracy as high
as 3--5\%. We took this into account in our spectral
analysis by adding a systematic error of 5\%. All of
the errors given in this paper are purely statistical.
The standard OSA software of version 4.1, which was
provided by the INTEGRAL Science Data Center
(http://isdc.unige.ch), was used for our timing analysis 
of the source on pulsation period time scales and
for our analysis of the JEM-X data.
 
   Figure 1 shows the IBIS 20--60 keV map of
the LMC region. This map was constructed over
the entire period of the source's observations in
January 2003. Apart from the pulsar LMC X-4
itself, the black hole LMC X-1 and the single pulsar
PSR 0540­69 are detected in the image at a statistically 
significant level. No emission is detected from
the SN 1987A remnant (Shtykovskiy et al. 2005).

\newpage

\section*{TIMING ANALYSIS}

\centerline{\it The Precession Period}

    As was said in the Introduction, there is a period
of about 30 days in the binary LMC X-4 that is
most likely related to the eclipse of the neutron star's
emitting regions by a precessing tilted accretion disk.
To determine the exact precession period, we used
the ASM/RXTE data obtained from January 1996
through July 2004. The best precession period of
the accretion disk was found by the epoch-folding
technique (Leahy et al. 1983) to be 30.275 days. We
used the following method to estimate the error of
this measurement: for each point on the light curve of
the source within a $1\sigma$ error in the flux, we randomly
chose a point that was used as a new flux from the
source. In this way, we modeled another light curve
for which, in turn, the best period was determined. We
obtained a total of about 100 periods whose spread
was considered to be a $1\sigma$ confidence interval. Such
an analysis yielded the following mean precession
period of the accretion disk for 1996--2004: 
$P_{prec}=30.275\pm0.004$ days.

    The hard X-ray (20--50 keV) light curve of the
source shown in Fig. 2a has a characteristic shape.
About 2 days after the beginning of its observations, 
the pulsar switched to the on state; in another
5 days, its flux increased by a factor of about 6--8 
and reached a peak ($\sim 70$ mCrab). The arrows
in the figure indicate the switch-on and switch-off
times estimated more than 20 years ago by Lang
et al. (1981); the switch-on time was calculated by
using the ephemerides derived by these authors; the
switch-off time was determined by assuming that the
high state of the source lasted  $\sim60$\% of the entire
precession period (Lang et al. 1981). We see that
the presumed zero phase (switch-on time) of the
superorbital cycle is greatly shifted relative to the
current INTEGRAL measurements. Note that the
same relationship between the durations of the high
and low states is also obtained from the INTEGRAL
observations of the source (Fig. 2a).

    Unfortunately, the switch-on time of the object
under study was not observed by the INTEGRAL
observatory directly, and we had to use interpolation
to determine the zero phase of the precession period.
The switch-on time of the pulsar LMC X-4 estimated 
in this way is MJD  $52644.5\pm1.0$. Owing to
good statistics and a long series of observations, we
were able to determine the mean precession period
with a high accuracy (see above), which ensures the
determination of the presumed zero phase 20 years
ago with an accuracy of  $\sim 1$ day by assuming the
period to be constant. Using our value of the precession 
period, we can calculate the presumed zero
phase closest to that measured by Lang et al. (1981).
Our result is either  $\sim18.5$ days before or  $\sim12$ days
after this time. Thus, the precession period of the
accretion disk, on average, must either increase with
$\dot{P}\sim1.3\times10^{-3}$ s s$^{-1}$ or decrease with 
$\dot{P}\sim-2\times10^{-3}$ s s$^{-1}$. Here, we give lower limits on the rate of
change in the period, since we disregard the change
that is a multiple of an integer number of periods.
 
   On the other hand, the mean rate of change in
the precession period can be estimated by using
our values and those obtained by Lang et al. (1981).
This estimation yields a value of 
$\dot{P}=(-2.5\pm0.8)\times10^{-5}$ s s$^{-1}$, which is much lower than our previous 
estimate obtained by extrapolating the switch-on 
times and agrees, within the error limits, with the
value determined by Paul and Kitamoto (2002).

    The observed discrepancy in the estimated rates
of change in the precession period may stem from the
fact that the precession period of the accretion disk
for the pulsar LMC X-4 has evolved not uniformly,
but with spinups/spindowns much higher than the
average level.

    Using long-term ASM/RXTE observations (January 
1996--August 2002), Klarkson et al. (2003)
constructed a dynamic power spectrum for the source
under study. Throughout the period of observations,
the precession period was constant near 30.28 days.
However, due to the $3\sigma$ error of $\pm0.46$ days, no firm
conclusion about the precession-period variations
can be reached.

   Figure 2b shows the averaged light curve of the pulsar
LMC X-4 constructed from ASM/RXTE data over
$\sim8.5$ years of its observations and folded with the
best period of 30.275 days. Interestingly, there is a
statistically significant rise in the source's intensity
with a duration of about three days near phase $\sim0.8$
of this cycle. A similar feature in the light curve of
the pulsar can be seen in Figs. 1 and 3 from Lang
et al. (1981). A similar rise in the source's intensity
during the off state is also observed in another X-ray
pulsar, Her X-1 (Jones and Forman 1976).

\centerline{\it The Orbital Period}

    Figure 3a shows the 20--50 keV light curve for the
pulsar LMC X-4 when the source was near the peak
of its observed X-ray flux. The times of the eclipses of
the X-ray source by its optical companion are clearly
seen from this figure. The ingress and egress were
accompanied by smooth variations in the pulsar's
observed intensity. However, a significant rise in the
flux from the source with a duration of about three
hours was observed a short time (of the order of an
hour) after the egress. The ingress and egress duration 
can be estimated from the source's light curve,
$\sim3$ ks. At the neutron star's orbital velocity of about
500 km s$^{-1}$ (Kelley et al. 1983), this corresponds to
the transition region's size of  $\sim1.5\times10^6$ km, which
is comparable to the sizes of the atmospheres of late-O 
main-sequence stars (Woo et al. 1996).
 
   The time dependence of the hardness, which is
defined as the ratio of the fluxes from the source in
the 40--60 and 20--40 keV energy bands, is shown
in Fig. 3b. We see a tendency for the radiation from
the object under study to soften as it approaches
an eclipse, but this dependence breaks down immediately 
after the eclipse. The hardness during the
eclipses themselves is not shown in the figure, because 
no statistically significant flux can be recorded
from the source.

   The dashed lines indicate the eclipse ephemerides
taken from the papers by Li et al. (1978), Lang et
al. (1981), and Levine et al. (2000). Note that the
orbital period in the binary is highly stable and has
been almost constant over more than twenty years of
the pulsar's studies. The INTEGRAL observations
of the source in 2003 confirm this conclusion (see
Fig. 3). Several authors (Woo et al. 1996; Levine
et al. 2000) have attempted to measure the rate of
change in the orbital period, but the significance of the
values obtained is low 
($\dot{P}_{orb}/P_{orb}=(-5.3\pm2.7) \times 10^{-7}$ yr$^{-1}$ and
$\dot{P}_{orb}/P_{orb}=(1.1\pm0.8) \times 10^{-6}$ yr$^{-1}$,
respectively) and is consistent with the observational
data.

   Note that the low-state flux from the source does
not drop to zero and is recorded by the INTEGRAL
instruments, although it decreases. Thus, for example, 
according to the observational data spanning the
time interval MJD 52662.5--52665, the 20--50 keV
flux from the source was $2.6\pm0.4$ mCrab, while several 
days later, MJD 52667--52668, it decreased by
almost a factor of 2, $1.3\pm0.5$ mCrab.

    The intensity of the pulsar LMC X-4 decreases
sharply during its X-ray eclipses. We analyzed the
flux during eclipses for two time intervals, when the
source was in its high and low states (MJD 52648.8
and MJD 52665, respectively). In both cases, no
emission was detected from the object under study at
a statistically significant level, and we obtained only
upper limits on its 20--50 keV flux, which proved to
be identical,  $\sim1.2$ mCrab ($1\sigma$). Note that during the
X-ray eclipse on January 20, 2003, (MJD 52659.94--52660.17), 
when the source's intensity decreased, its
observed 20--50 keV flux was significantly higher
than the above upper limits, $3.7\pm1.0$ mCrab, but
the significance of this measurement is low. A similar
analysis for the soft (0.1--2.4 keV) X-ray emission
was performed by Woo et al. (1995). The results
obtained by these authors suggest that the source's
intensity is constant during its eclipses, which they
interpreted as evidence that the pulsar's intrinsic luminosity 
is constant. This conclusion was based on
the assumption that the X rays observed during an
eclipse are scattered by the widely distributed matter
around the source, possibly the corona above the
outer edges of the accretion disk, mush as is the case
in the pulsar Her X-1 (Lutovinov et al. 2000).

\centerline{\it Flares}

    Many authors have described the flaring activity of
the pulsar LMC X-4. Short flares (up to 1000 s in duration), 
during which the pulsar's intensity changes
by a factor of 2 to 5, are most characteristic of the
source under study (Levine et al. 2000).

    An rise in the observed flux from the source was
recorded during its low state (MJD 52666) on a time
scale atypical of it. The flare duration was about 10 h
(orbital phase 0.3--0.6), and the peak intensity ($\sim10$
mCrab) exceeded the average low-state level by
a factor of about 4. Figure 4 shows the flare profile
constructed from IBIS data in the 20--50 keV energy
band.

\section*{SPECTRAL ANALYSIS}

    In the previous section, we showed that a significant 
number of features could be distinguished in
the light curve of the pulsar LMC X-4. Therefore,
studying the spectral evolution of the source with time
is of considerable interest. Another goal of the spectral 
analysis is the search for the possible cyclotron
resonance absorption line whose existence was mentioned 
by several authors (see the Introduction).

    A power law with an exponential high-energy cut-off 
(White et al. 1983), which is typical of this class
of objects, was chosen as a basis for the spectral
analysis of the source over a wide (4--100 keV) energy
range:

\begin{equation}
I(E)=A\,E^{-\alpha} \times  
\left\{\begin{array}{c} 1, 
\mbox{ $E<E_{c}$};\\ 
\exp{[-(E-E_{c})/E_{f}\,]}, 
\mbox{}\ E\geq E_{c},\\
\end{array}\right.
\end{equation} 

where $E$ is the photon energy in keV, $A$ is the 
normalization of the power-law component, $\alpha$ is the photon
spectral index, $E_{c}$ is the cutoff energy, and $E_{f}$ is the 
$e$-folding energy in the source's spectrum. 
We identified
a total of four states in which an independent analysis
of the pulsar's spectrum was performed: the high
state (without including the X-ray eclipse times), the
bursts immediately after the source's egress, the low
state (without including the X-ray eclipse times), and
the flare occurred on January 27, 2003 (MJD 52666).
 
   The best-fit parameters for the source's spectra
based on model (1) are given in the table. We were
able to reconstruct the spectra over a wide energy
range (using JEM-X data) only in the high state.
Since the JEM-X monitor is not sensitive enough,
the source was detected at a statistically significant
level only by the IBIS telescope. Therefore, when fitting 
the spectra in this state, we fixed the parameters
in the standard X-ray energy range ($<20$ keV) at
the values obtained for the high state. We see from
the table that the shape of the source's spectrum and
its parameters (slope and cutoff parameters) remain
almost constant with the pulsar's intensity variations.
Figure 5 shows the energy spectrum of LMC X-4
reconstructed from INTEGRAL data in a wide energy
range for the high state.
  
  To test the hypothesis about the presence of a
cyclotron feature in the source's spectrum within
the energy range 4--100 keV, we modified the fitting
model by adding the corresponding component:

\begin{equation}
exp\left[- \frac{A_{cyc}W_{cyc}^2 (E/E_{cyc})^2}{(E-E_{cyc})^2+W_{cyc}^2}\right]\
,\end{equation}

where $E$ is the photon energy in keV, $E_{cyc}$ is the
cyclotron energy, $W_{cyc}$ is the width of the cyclotron
line, and $A_{cyc}$ is its depth.

    Using the modified model, we fitted the pulsar's
spectrum during the high state, except the X-ray
eclipse times. We used the same procedure that was
employed to study the X-ray pulsar KS 1947+300
(Tsygankov and Lutovinov 2005): the energy of the
center of the presumed cyclotron line $E_{cyc}$ was varied
over the range 5--100 keV at 5-keV step, while its
width was fixed at 5 keV; for each trial energy of the
line center, we tested the significance of an improvement 
in statistics using the $\Delta\chi^2$ test. As a result,
we found no such energy of the cyclotron line $E_{cyc}$ in
the energy range under study whose inclusion in the
model would lead to an improvement in the quality of
the spectral fit by more than $2\sigma$.
 
   The result obtained may suggest that either the
energy of the cyclotron line lies outside our energy
range (4--100 keV) or the INTEGRAL instruments
are not sensitive enough for the cyclotron line to be
detected in the spectrum of the source LMC X-4. The
relationship between the spin period of the neutron
star and its luminosity, assuming that the period is
close to its equilibrium value, argues for the hypothesis 
of a strong magnetic field ($>10^{13}$ G) in the binary
(Woo et al. 1996).

\section*{CONCLUSIONS}

    We presented the results of our long-term timing
and spectral analyses for the X-ray pulsar LMC X-4 
performed by using the INTEGRAL observations
in January 2003 and the long-term (1996--2004)
ASM/RXTE observations. We determined the precession 
period ($30.275\pm0.004$ days) averaged over
the last  $\sim8.5$ years with a high accuracy. Since the
source's switch-on times predicted by different authors 
do not coincide, we showed that this parameter
is not constant and most likely varies nonuniformly.
The IBIS/INTEGRAL data revealed small bursts
that emerged after the egress of the X-ray source.
We detected a flare about 10 h in duration during the
low state. On the time scales of the orbital period, its
value determined more than 20 years ago is stable and
satisfies well our observational data. We performed a
spectral analysis for different states distinguishable
by the object's intensity. The spectrum of the source
is described by a power law with a high-energy cutoff,
which is characteristic of this class of objects. To
test the hypothesis about the presence of a cyclotron
feature in the source's spectrum within the energy
range 4--100 keV, we properly modified the fitting
model. This analysis showed that there is no such
feature in the source's spectrum at a confidence
level exceeding $2\sigma$. A comparison of the derived
constraints with the results of other authors (Woo et
al. 1996) is more likely indicative of a strong magnetic
field ($>10^{13}$ G) on the neutron-star surface than a
weak field ($<5\times10^{11}$ G).

\section*{ACKNOWLEDGMENTS}

   We wish to thank E.M. Churazov, who developed
the methods and software for analyzing the data from
the IBIS telescope of the INTEGRAL observatory.
This work was supported by the Russian Foundation
for Basic Research (project no. 04-02-17276). We
used the data retrieved from the High-Energy Astrophysics 
Archive at the Goddard Space Flight Center
of NASA and the data retrieved from the Archive
of the INTEGRAL Science Data Center (Versoix,
Switzerland) and the Russian INTEGRAL Science
Data Center (Moscow, Russia). This study was performed 
in part during the visits to the INTEGRAL
Science Data Center (Versoix, Switzerland), and we
are grateful to its staff for hospitality. A.A. Lutovinov
is also grateful to the ESA for financial support of
some of these visits.

 \pagebreak
%****************************************************************

\section*{REFERENCES}

 1. C. Chevalier and S. Ilovaisky, Astron. Astrophys. {\bf 59},
    L9 (1977).

2. Clarkson W.I., Charles P.A., Coe M.I., et al., Mon. Not. Roy. Astron. Soc. {\bf
343}, 1213 (2003)

 3. N. A. Eismont, A. V. Ditrikh, G. Janin, et al., Astron.
    Astrophys. {\bf 411}, L37 (2003).

 4. A. Epstein, J. Delvaille, H. Helmken, et al., Astrophys. J. {\bf 216}, 103 (1977).

 5. P. Ghosh and F. Lamb, Astrophys. J. 234, 296 (1979).
 
6. C. Jones and W. Forman, Astrophys. J. {\bf 209}, L131
    (1976).

 7. R. L. Kelley, J. G. Jernigan, A. Levine, et al., Astrophys. J. {\bf 264}, 568 (1983).

 8. A. La Barbera, L. Burderi, T. Di Salvo, et al., Astrophys. J. {\bf 553}, 375 (2001).

 9. F. L. Lang, A. M. Levine, M. Bautz, et al., Astrophys.
    J. {\bf 246}, L21 (1981).

 10. D. A. Leahy, R. F. Elsner, and M. C. Weisskopf,
    Astrophys. J. {\bf 272}, 256 (1983).

11. F. Lebrun, J. P. Leray, P. Lavo cat, et al., Astron. Astrophys. {\bf 411}, L141
(2003)

12. A. Levine, S. Rappaport, A. Putney, et al., Astrophys.
    J. {\bf 381}, 101 (1991).

13. A. Levine, S. Rappaport, and G. Zojcheski, Astrophys. J. {\bf 541}, 194 (2000).

14. F. Li, S. Rappaport, and A. Epstein, Nature {\bf 271}, 37
    (1978).

15. N. Lund, S. Brandt, C. Budtz-Jo ergesen, et al., Astron. Astrophys. {\bf 411}, L231 (2003)

16. A. A. Lutovinov, S. A. Grebenev, M. N. Pavlinsky, and
    R. A. Sunyaev, Astron. Lett. 26, 691 (2000).

17. A. A. Lutovinov, S. V. Mol'kov, and M. G. Revnivtsev,
    Pis'ma Astron. Zh. {\bf 29}, 803 (2003) [Astron. Lett. 29,
    713 (2003)].

18. A. A. Lutovinov, S. S. Tsygankov, M. G. Revnivtsev,
    et al., in Proceedings of the V INTEGRAL Workshop, ESA SP, {\bf 552}, 253 (2004).

19. T. Mihara, PhD Thesis (Univ. Tokyo, Tokyo, 1995).

20. S. Naik and B. Paul, Astron. Astrophys. {\bf 401}, 265
    (2003).

21. B. Paul and S. Kitamoto, J. Astrophys. Astron. {\bf 23}, 33
    (2002).

 22. M. G. Revnivstev, R. A. Sunyaev, D. A. Varshalovich, et al., Pis'ma Astron. Zh. {\bf 30},
430 (2004) [Astron. Lett. {\bf 30}, 382 (2004)]

23. G. K. Skinner, S. Shulman, and G. Share, Astrophys.
    J. {\bf 240}, 619 (1980).

24. H. Tananbaum, H. Gursky, E. Kellog, et al., Astrophys. J. {\bf 174}, L143 (1972).

25. P. E. Shtykovskiy, A. A. Lutovinov, M. R. Gilfanov,
    and R. A. Sunyaev, Pis'ma Astron. Zh. {\bf 31}, 284 (2005)
    [Astron. Lett. 31, (2005)].

26. S. S. Tsygankov and A. A. Lutovinov, Pis'ma Astron.
    Zh. {\bf 31}, 99 (2005) [Astron. Lett. 31, 88 (2005)].

27. P. Ubertini, F. Lebrun, G. Di Co cco, et al., Astron. Astrophys. {\bf 411}, L131 (2003)

28. N. White, J. Swank, and S. Holt Astrophys. J. {\bf 270},
    771 (1983).

29. C. Winkler, T. J.-L. Courvoisier, G. Di Co cco, et al., Astron. Astrophys. {\bf
411}, L1 (2003)

30. J. W. Woo, G. W. Clark, and A. M. Levine, Astrophys.
    J. {\bf 449}, 880 (1995).

31. J. W. Woo, G. W. Clark, A. M. Levine, et al.), Astrophys. J. {\bf 467}, 811 (1996).

\pagebreak
%************************************************************************

\begin{landscape}
\begin{table}[h]
\centering

\hspace{-3mm}{Best-fit parameters for the spectrum of LMC X-4 derived from INTEGRAL data (JEM-X + IBIS)\a}

\vspace{2mm}

\hspace{-5mm}\begin{tabular}{@{}c|c|c|c|c|c}
\hline\hline
&&&&\\ [-4mm]
State & Luminosity,& $\alpha$ & $E_{c}$, & $E_{f}$, &$\chi^{2}_{\,N} (N)$\c \\
&$\times10^{37}$~ erg s$^{-1}$ \b&&keV&keV&\\[2mm]

\hline
&&&&&\\ [-4mm]
High (MJD 52646-52650) &38.7&$0.20\pm0.15$&$9.1\pm0.8$&$11.0\pm0.6$&0.88(128)\\[2mm]
Flares (in high state)&47.9&$0.44\pm0.20$&$8.9\pm1.0$&$11.9\pm0.8$&0.82(126)\\[2mm]
Low (MJD 52662.5-52668)&1.4&$0.2$\d&$9.1$\d&$14.8\pm3.2$&1.19(4)\\[2mm]
Flare (MJD 52666) &3.1&$0.2$\d&$9.1$\d&$17.1\pm5.3$&0.41(6)\\[2mm]
\hline
\end{tabular}

\vspace{3mm}

\begin{tabular}{cl}
\a&  All errors are given at a 1$\sigma$ level.\\
\b& The 4--100 keV luminosity at an assumed distance to the source of 50 kpc.\\
\c& The $\chi2$ value normalized to the number of degrees of freedom $N$.\\
\d& The parameters were fixed at the values obtained for the high state.\\

\end{tabular}
\end{table}
\end{landscape} 

\pagebreak

%****************************************************************

\begin{figure*}[t]
\centerline{\includegraphics[width=17cm,bb=40 200 560 590,clip]{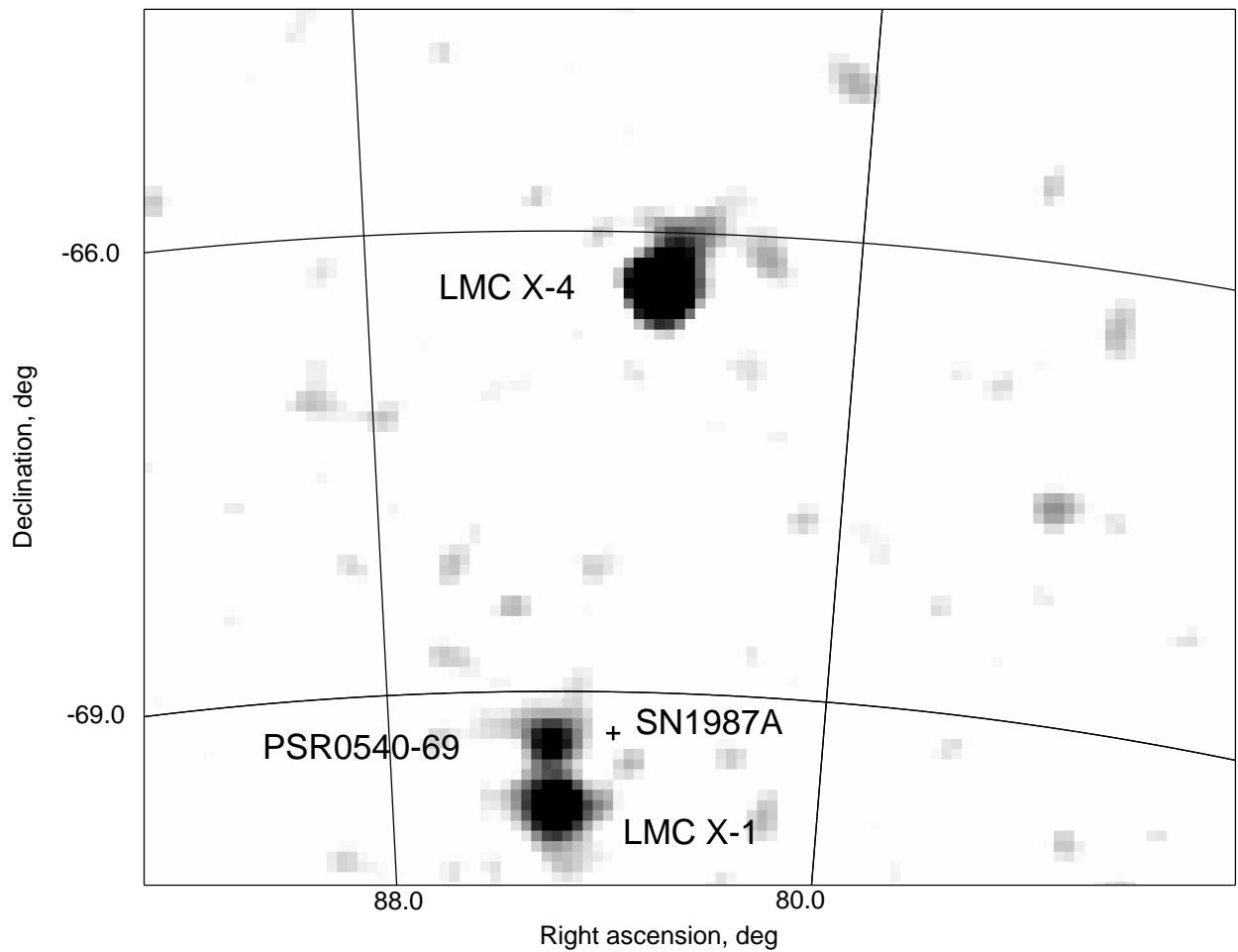}}

\vfill

\caption{IBIS/INTEGRAL 20--60 keV image of the LMC region containing 
the X-ray pulsar LMC X-4. The sources detected
at a statistically significant level are labeled; the position of the SN 1987A 
remnant is indicated by the cross.
  }
\end{figure*}
\centerline{  }

\newpage

\begin{figure*}[t]
\centerline{\includegraphics[width=17cm,bb=20 190 565 700,clip]{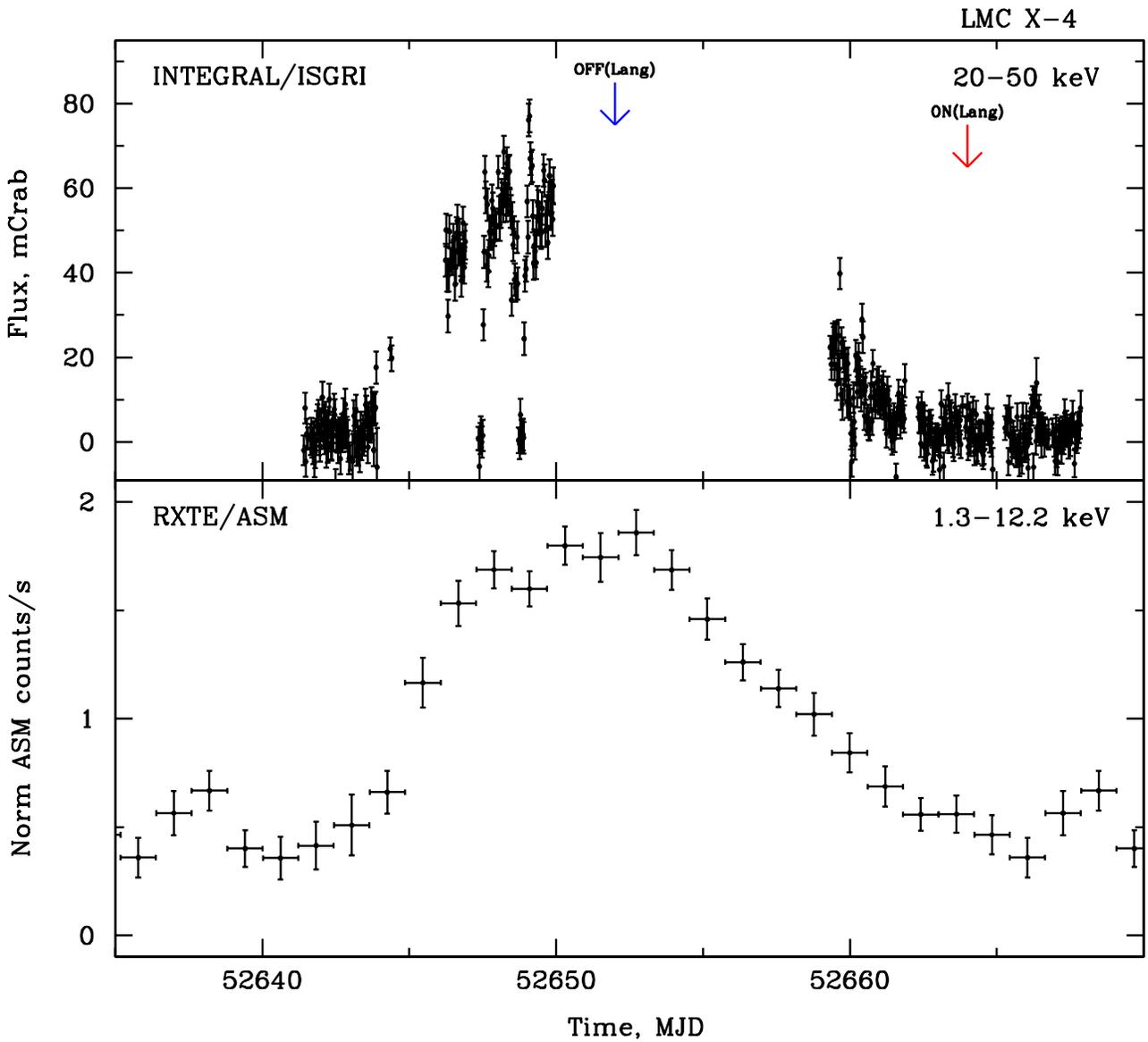}}

\vfill

\caption{(a) Hard X-ray (20--50 keV) light curve for the pulsar LMC X-4 constructed from IBIS/INTEGRAL data. The arrows
indicate the switch-on and switch-off times for the source, as estimated by other authors. (b) The light curve for the source
constructed from ASM/RXTE data over eight years of observations and folded with the best period derived from these data.
The errors correspond to one standard deviation.}

\end{figure*}
\centerline{  }

\newpage

\begin{figure*}[t]
\includegraphics[width=17cm,bb=15 190 570 690,clip]{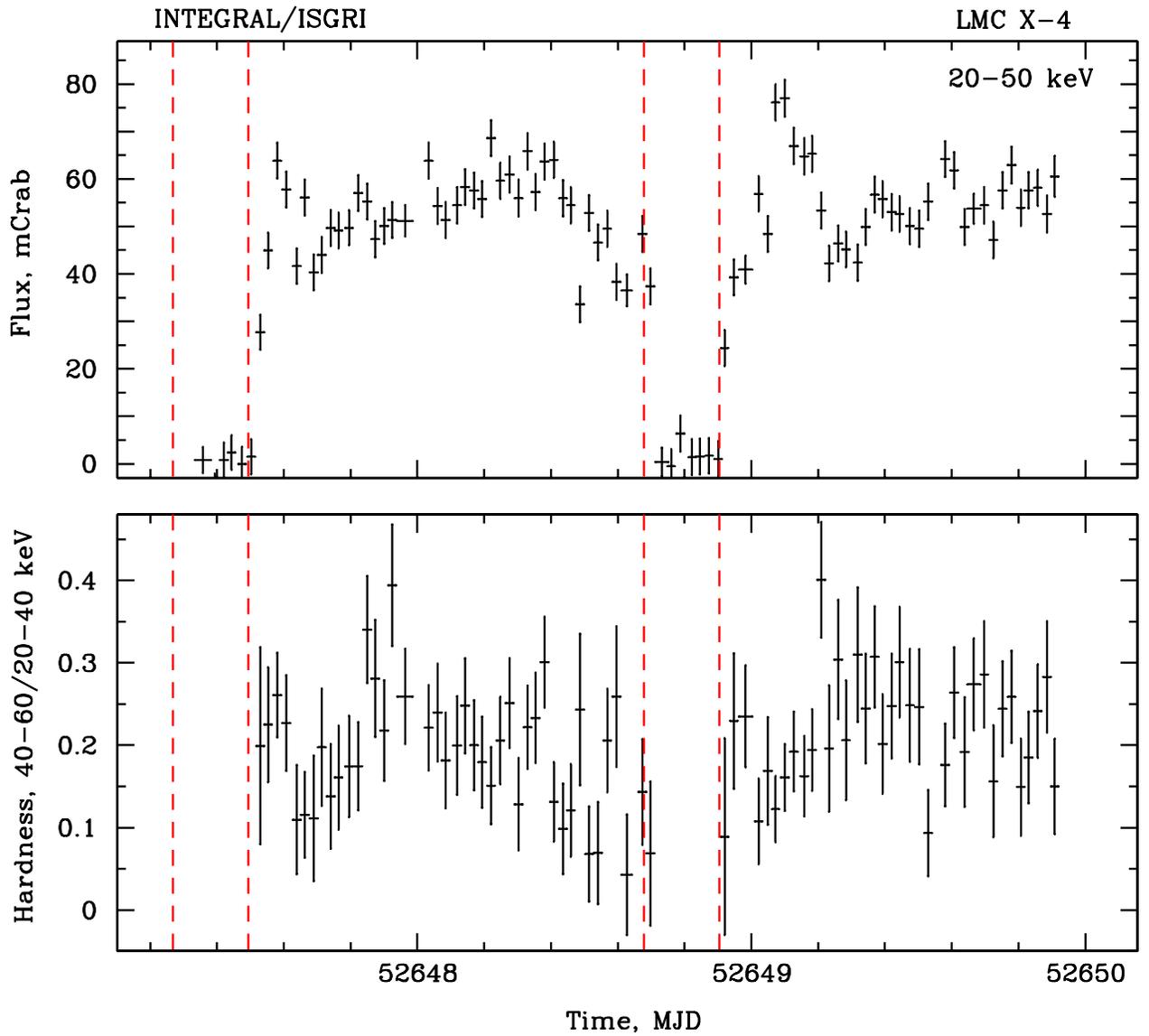}

\vfill

\caption{(a) Orbital light curve for the pulsar LMC X-4 constructed from IBIS data during its high state. The dashed lines
indicate the ingress and egress times estimated by Lang et al. (1981). (b) Variations of the source's hardness on the time scale of the
orbital period. The errors correspond to one standard deviation.}

\end{figure*}
\centerline{  }

\newpage

\begin{figure*}[t]
\includegraphics[width=17cm,bb=18 415 460 690,clip]{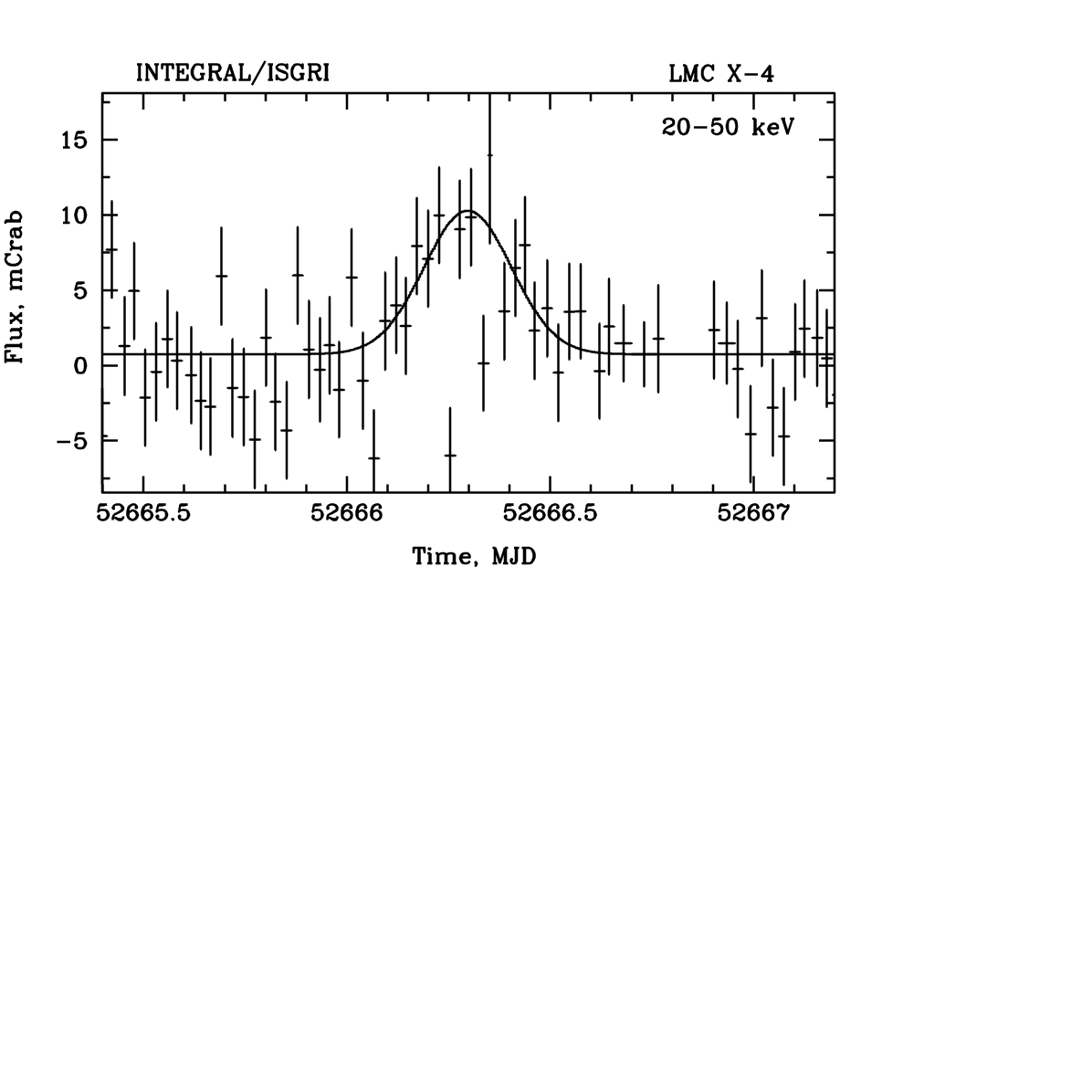}

\vfill

\caption{Flare detected by the IBIS telescope of the INTEGRAL observatory from the pulsar LMC X-4 on January 27, 2003,
during the low state. The solid line indicates a Gaussian fit to the flare profile.}

\end{figure*}
\centerline{  }

\newpage

\begin{figure*}[t]
\vbox{
\centerline{\includegraphics[width=17cm,bb=35 360 540 700,clip]{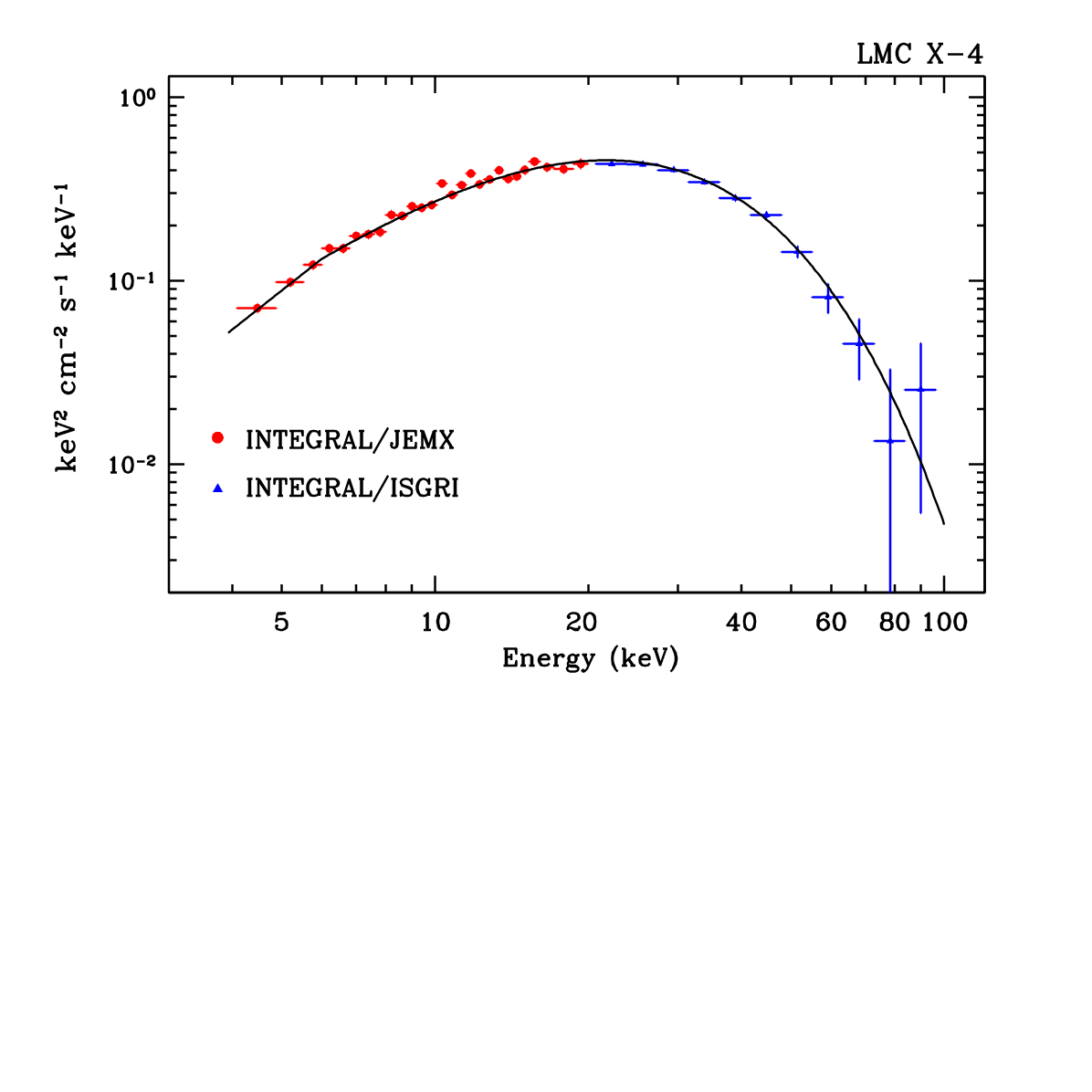}}
}
\vfill

\caption{Energy spectrum of the source LMC X-4 during its high state constructed from JEM-X and IBIS data. The dots
indicate the experimental spectrum; the solid line represents its power-law fit with a high-energy cutoff.}

\end{figure*}
\centerline{  }

\end{document}